\documentclass[twocolumn, amsmath, amssymb, prl, superscriptaddress]{revtex4}

\usepackage{graphicx}
\usepackage{dcolumn}
\usepackage{bm}

\usepackage{amssymb}
\usepackage{amsmath} 

\newcommand\ee{\end{equation}}
\newcommand\be{\begin{equation}}
\newcommand\eea{\end{eqnarray}}
\newcommand\bea{\begin{eqnarray}}
\newcommand{\sfrac}[2]{{\textstyle\frac{#1}{#2}}}
\newcommand\di{\partial}

\begin{document}


\title{Derrick's theorem beyond a potential}

\author{Solomon Endlich}
\affiliation{%
Physics Department and
Institute for Strings, Cosmology and Astroparticle Physics,\\
Columbia University, New York, NY 10027, USA
}%

\author{Kurt Hinterbichler}
\affiliation{%
Center for Particle Cosmology, University of Pennsylvania, \\ Philadelphia, PA 19104, USA
}%
\affiliation{%
Physics Department and
Institute for Strings, Cosmology and Astroparticle Physics,\\
Columbia University, New York, NY 10027, USA
}%

\author{Lam Hui}
\affiliation{%
Physics Department and
Institute for Strings, Cosmology and Astroparticle Physics,\\
Columbia University, New York, NY 10027, USA
}%

\author{Alberto Nicolis}
\affiliation{%
Physics Department and
Institute for Strings, Cosmology and Astroparticle Physics,\\
Columbia University, New York, NY 10027, USA
}%

\author{Junpu Wang}
\affiliation{%
Physics Department and
Institute for Strings, Cosmology and Astroparticle Physics,\\
Columbia University, New York, NY 10027, USA
}%


\begin{abstract}
Scalar field theories with derivative interactions are known to possess 
solitonic excitations, but such solitons are generally unsatisfactory because 
the effective theory fails precisely where nonlinearities responsible 
for the solitons are important. A new class of theories possessing (internal) galilean invariance can in principle
bypass this difficulty. Here, we show that these galileon theories
do not possess stable solitonic solutions. As a by-product, we show that
no stable solitons exist for a different class of derivatively coupled theories, 
describing for instance the infrared dynamics of superfluids, fluids, solids and some k-essence models.
\end{abstract}

\pacs{}
\maketitle










%



A famous theorem by Derrick states that in a theory of scalar fields with potential interactions only, there cannot be stable soliton solutions in more than one spatial dimension \cite{D}. By `soliton' it is usually meant a static non-trivial solution of the field equations with finite total energy, and in the present letter we stick to that definition. It is also understood that there is no external source sustaining the field profile -- the  solution should be sustained by non-linearities in the fields.
Briefly speaking, Derrick's proof proceeds by contradiction: 
suppose a solitonic solution $\phi_0(\vec x)$ exists, then consider
deformations $\phi_\lambda(\vec x) = \phi_0(\lambda \vec x)$ labeled by
a dilation parameter $\lambda$, and show that
in more than one spatial dimension no (stable) stationary point of energy
exists with respect to $\lambda$ for a scalar with purely potential
interactions.

What about interactions that are not purely of the potential form?  After all, derivative interactions are generically expected in any low-energy effective theory, and beyond the two-derivative level one indeed introduces new powers of $\lambda$ into Derrick's argument, thus in principle allowing for more stationary points. The problem, however, is that such interactions are always non-renormalizable. Therefore, either field gradients are so mild that they are negligible, or if they become important, the derivative expansion is expected to break down. In such a regime the effective theory is not enough to ascertain the existence of solitons, and one needs a UV completion (which comes with new or different degrees of freedom -- thus in a sense redefining the problem we set out to solve). For example,
this is the case for skyrmions in the chiral Lagrangian \cite{skyrmion} -- to evade Derrick's theorem one needs to rely on terms beyond the two-derivative level, and there is no reason why further terms in the derivative expansion should be negligible.


There is, however, a class of derivatively coupled theories where these conclusions do not necessarily apply. 
Consider the theory of a Goldstone boson $\pi$ invariant under constant shifts of $\pi$ and its first {\em derivatives}
\be
\pi (x) \to \pi + c + b_\mu x^\mu \; .
\ee
Such a symmetry has been dubbed `galilean invariance', and $\pi$ the `galileon' \cite{NRT}. Galilean invariance forces the equations of motion to involve {\em at least} two derivatives acting on each field. Absence of ghosts in a non-linear regime demands that there be {\em at most} two derivatives on each field. Therefore, in order to have a galilean invariant dynamics that we can trust -- at least classically -- even when non-linearities are important, we need precisely two derivatives per $\pi$ in the field equations. This requirement corresponds to having Lagrangian terms of the form
\be \label{Ln}
{\cal L}_n \sim \di \pi \, \di \pi\, (\di^2 \pi)^{n-2}
\ee
with suitable Lorentz contractions (and dimensionful coefficients). Such combinations have been thoroughly classified: in $d+1$ spacetime dimensions there are $d+1$, corresponding to $n= 2, \dots, d+2$ \cite{NRT}.  The $n=2$ term is just the usual kinetic term $(\di \pi)^2$. That with $n=3$ is $(\di \pi)^2 \Box \pi$, and so on. 
It is worth mentioning that 
the galileon describes the sub-horizon dynamics of the scalar sector of the DGP \cite{DGP} model ($n=3$), and 
of more generic theories that modify general relativity in the IR \cite{LPR, NR, NRT}.

Of course classical consistency (e.g.~absence of ghosts) of non-linear solutions is only a necessary condition for these to be inside the effective theory. One also needs quantum effects to be small. Here we are rescued by galilean invariance, thanks to which the ${\cal L}_n$'s above do not get renormalized upon loop corrections, and terms with fewer derivatives per field are not generated quantum-mechanically \cite{LPR}.
Indeed the structure of divergences in the one-loop effective action (in $3+1$ dimensions) is schematically
\cite{NR}
\be
\Gamma^{\rm 1-loop} \sim \sum_m \big[ \Lambda^4 + \Lambda^2 \di^2 +
\di^4 \log \di^2 / \Lambda^2  \big] \bigg( \frac{\di \di \pi}{\Lambda^3} \bigg)^m  \; ,
\ee
where  $\Lambda$ is the mass-scale suppressing the galilean interactions (\ref{Ln}).  For simplicity, we cut off the UV divergences at $\Lambda$.  The sum runs over external legs. 
Now, $\alpha_{ cl} \equiv \di \di \pi / \Lambda^3$ is a measure of classical non-linearities. For instance the $n$-th order galilean interaction ${\cal L}_n $ is roughly $\alpha_{cl}^{n-2}$ times the kinetic energy for $\pi$. On the other hand the quantity suppressing quantum effects is really $\alpha_q \equiv \di^2/\Lambda^2$, in the sense that, 
factoring out two powers of $\pi$, we have
\be
\Gamma^{\rm 1-loop} \sim  \sum_{m'} \big[\alpha_q + \alpha_q^2 +
\alpha_q^3 \log \alpha_q  \big] \di \pi \di \pi \bigg( \frac{\di \di \pi}{\Lambda^3} \bigg)^{m'}  \; .
\ee
Even for non-linearities of order one, $\di \di \pi / \Lambda^3 \sim 1$, this is suppressed w.r.t.~to the tree-level interactions (\ref{Ln}) by explicit powers of $\alpha_q$
\footnote{Strictly speaking, calculable (and measurable) quantum effects are only those associated with log-divergences -- the last term in brackets. Power-divergences are instead completely regularization dependent, and indeed  in dimensional regularization they do not even show up. However here we also want to stress that our galilean Lagrangian does not even suffer from fine-tuning problems, which are associated with power divergences.}. 
This means that
for classical solutions with large non-linearities, quantum effects are small as long as gradients are mild $\di  \ll \Lambda$

These facts make it consistent (if not necessarily `natural') to postulate that the regime
of large {classical} non-linearities is within  the effective theory. The handful of galilean-invariant terms described above represent a consistent truncation of the theory in which we can self-consistently study non-linear classical solutions \cite{NR}.
It thus makes sense to ask whether there can be soliton solutions for the galileon.

There is a final subtlety we need to address. As we have seen, in order to have small quantum effects we need mild gradients, $\di \ll \Lambda$. However, if the only scale appearing in the Lagrangian is $\Lambda$ itself, and we introduce no external sources, it is clear that if a soliton exists it will have a size of order $\Lambda^{-1}$ and gradients of order $\Lambda$. Nevertheless, we can consider the situation where different galilean interactions, say ${\cal L}_3$ and ${\cal L}_4$, are weighed by parametrically different scales, say $\Lambda_3 \ll \Lambda_4$. If the soliton is sustained by a balance between these interactions, its size will be a see-saw combination of the two scales $\Lambda_3^{-1}$ and $\Lambda_4^{-1}$, and potentially parametrically bigger than either. On the other hand all our estimates above for $\Gamma^{\rm 1-loop}$ apply upon replacing $\Lambda$ with the lower between the two scales, $\Lambda_3$. In this example it is thus possible that there are classical soliton solutions that are within the effective theory. Notice that the `unnatural' tuning $\Lambda_3 \ll \Lambda_4$ is in fact technically natural, in that, as we mentioned, galilean interactions are not renormalized upon quantum corrections \footnote{It is worth mentioning that
a `standard' UV-completion -- i.e.~a relativistic UV completion
obeying the standard analyticity properties of S-matrix theory,
like renormalizable Lorentz-invariant QFTs and weakly coupled string
theories -- does not appear to exist for galilean theories \cite{AADNR, NRT2}.}.
 


Let us now return to our main goal of discovering solitons, or lack
thereof, in these theories. Suppose a solitonic solution $\pi_0 (\vec x)$
exists. Small perturbations $\varphi$ around it
lead to energy fluctuations of this form:
\be \label{quadratic_energy}
\delta E =\sfrac12 \int  d^d x \,  Z^{ij} (\vec x)\, \di_i \varphi \, \di_j \varphi \; ,
\ee
where $Z^{ij}$ is a matrix built out of second derivatives of the background solution $\pi_0$ \cite{NR, NRT},
\be \label{Zijform}
Z^{ij} = c_2 \, \delta^{ij} + c_3 \big( \delta^{ij} \nabla^2 \pi_0 - \di^i \di^j \pi_0 \big) + \dots \; ,
\ee 
where the $c_n$'s are  the coefficients of the terms (\ref{Ln}) in the galileon action. This is highly non-trivial. In principle, given the structure of a generic galilean term (\ref{Ln}), one expects contributions to the quadratic action for small fluctuations of three different forms:
\bea \label{4derivative}
\di \pi_0 \di \pi_0 (\di^2 \pi_0)^{n - 4} \: \di^2 \varphi \di^2 \varphi  &&\\  
\di \pi_0 (\di^2 \pi_0)^{n - 3} \: \di^2 \varphi \di \varphi && \\ 
(\di^2 \pi_0)^{n - 2} \: \di \varphi \di \varphi && ,
\eea
with suitable index contractions. Only the third structure is of the form we like, eq.~(\ref{quadratic_energy}).
However the second structure can always be rewritten as the third by integration by parts, since the $\partial^2 \varphi \partial \varphi$ piece is a total derivative:
\be
\di_\mu \di_\nu \varphi \, \di_\alpha \varphi = \di_{(\mu }\big(\di_{\nu)} \varphi \di_\alpha \varphi \big) 
- \sfrac12 \di_\alpha \big(\di_\mu \varphi \di_\nu \varphi \big) \; .
\ee
As for the first structure, it cannot arise because the equations of motion are second order. Indeed
for a non-trivial Lagrangian term $A^{\mu\nu \alpha\beta} \, \di_\mu\di_\nu \varphi \, \di_\alpha \di_\beta \varphi $, there is an associated four-derivative contribution to the $\varphi$ e.o.m.~ $ 2 A^{\mu\nu \alpha\beta} \, \di_\mu\di_\nu \di_\alpha \di_\beta \varphi $, which vanishes only if the tensor $A$ does. Since the e.o.m.~for the full galileon field are of second order by construction, a term of the form (\ref{4derivative}) cannot appear in the quadratic action for fluctuations. 

The quadratic structure in eq. (\ref{quadratic_energy}) takes
on significance in light of another fact, namely
the existence of zero modes, i.e. small perturbations with $\delta E = 0$.
The simplest example is the translational mode, a perturbation
of the form:
\begin{eqnarray}
\label{translation}
\varphi_\epsilon = \vec \epsilon \cdot \nabla \pi_0 \; ,
\end{eqnarray}
where $\vec \epsilon$ is some infinitesimal vector. In other words,
spatially displacing a soliton does not change its energy.

Now, recall the quadratic structure for $\delta E$:
for the soliton to be stable (or marginally so),
it is crucial that $Z^{ij} (\vec x)$ be a positive semi-definite 
matrix everywhere in space.
At large distances from the `core' of our soliton, where $\pi_0$ becomes very small, $Z^{ij}$ is dominated by the zeroth order term, which is positive definite for $c_2 > 0$ (eq. (\ref{Zijform})).
But the existence of a zero-energy perturbation then implies that in some other region $Z^{ij}$ must develop a negative eigenvalue: since the integrand in eq.~(\ref{quadratic_energy}) is {\em strictly} positive at large $\vec x$, somewhere else it should become negative for the integral to yield zero.

The existence of negative eigenvalues for $Z^{ij}$ in some region means that suitably chosen localized
perturbations can be found that lower the energy of the solution -- it is enough to pick very short wavelength wave-packets with momentum along the negative eigenvalue direction \cite{NR}. Such an instability, which is associated with negative gradient energy for certain fluctuations, is much worse than an instability associated with a negative mass term.
This is because the former also plagues very short-wavelength fluctuations, down to the UV cutoff of the theory. The decay rate is thus extremely fast, dominated by the shortest scales in the theory, and cannot be reliably computed within the effective theory.   Conversely, a standard tachyon-like instability, like Jeans's or that associated with the negative mode of a true-vacuum bubble, is dominated by modes with momenta of order of the tachyonic mass scale, which can be parametrically smaller than the UV cutoff. As a consequence a tachyon-like instability can be slow, and its evolution can be consistently studied inside the effective theory, along with its interesting phenomenological consequences \footnote{Having a flat direction -- a trajectory
in field space along which $E$ remains constant --
is not by itself necessarily a sign of instability. 
In the case of a flat direction in a multi-field potential, there is a continuum of different configuration all with the same energy,  but to go from one to the other, one should change the boundary conditions at infinity. Instead, local excitations have positive-definite energy coming from the field gradients. 
So, in that case, the flat direction does not correspond to localized perturbations of the solution, and consequently each point along the flat direction corresponds to a stable solution.
However in our case the situation is very different.
The translation  zero-modes, eq. 
(\ref{translation}), are {\it local} perturbations
because $\pi_0$ is localized.}.

In conclusion, there is no consistent soliton solution in the galileon theory, in any number of spacetime dimensions.
Notice that our proof of instability crucially relies on the {\em purely kinetic} structure of the quadratic energy for small fluctuations, eq.~(\ref{quadratic_energy}). So, it does not apply to other theories that are known to possess soliton solutions, like e.g.~a scalar theory with a potential in $1+1$ dimensions. There, the soliton has zero-energy translational modes, yet the vanishing of their energy is accomplished by having a localized negative {\em mass term} for small fluctuations. As a consequence, if one tries to construct negative-energy fluctuations by localizing them where the mass term is negative, one in fact enhances the positive-definite kinetic energy -- which dominates over the mass term for large gradients -- thus ending up with positive overall energy. Indeed one can show that for `kinks' in $1+1$ dimensions the energy spectrum of small fluctuations in bounded by zero from below \cite{coleman}.

As an interesting aside, it is worth pointing out that were the galileon soliton
to exist, it would have zero total energy. To see this, consider
the galileon action which takes
the form $S = \int d^{d+1} x \, \sum_n c_n {\cal L}_n$ (eq. (\ref{Ln})).
The total energy of a static configuration is
\footnote{The fact that for static configurations the Hamiltonian density is minus the Lagrangian density may seem nontrivial here, due to the higher-derivative nature of the action. In fact the $T_{00}$ one gets by deriving the action w.r.t.~the metric is {\em not} $-{\cal L}$. On the other hand the $T_{00}$ derived {\em \`a la} Noether is always $-{\cal L}$ for static fields.  The difference between the two is a total spatial derivative, which integrates to zero for localized configurations.}
\be
E = - \int \! d^{d} x \, \sum_n c_n {\cal L}_n \equiv \sum_n  E_n \; ,
\ee
and the soliton solution $\pi_0(\vec x)$ we are looking for should be a local minimum of this.
Now we apply a minor generalization of Derrick's argument. Consider the field configuration obtained by rescaling the $\vec x$-dependence of $\pi_0$ as well as its overall normalization
\be \label{rescale}
\pi_{\lambda, \omega} (\vec x) \equiv \omega \, \pi_0(\lambda \,\vec x).
\ee
A necessary condition for $\pi_0$ to be a solution is that
\be
\di_\lambda E({\lambda, \omega}) \big|_{(1,1)} = 0 \; ,\qquad \di_\omega E({\lambda, \omega}) \big|_{(1,1)} = 0  \; ,
\ee
where $E({\lambda, \omega})$ is the energy of $\pi_{\lambda, \omega} $.
Now, the $n$-th order galilean invariant term (\ref{Ln}) involves $n$ fields and $2n-2$ derivatives. We thus have
\be \label{En}
E_n (\lambda, \omega ) = \lambda^{2n-2-d} \omega^n E_n^{(0)} =\frac{1}{\lambda^{d+2}} (\lambda^2 \omega)^n  E_n^{(0)}  \; ,
\ee
where $E_n^{(0)} $ is the $n$-th order term in the energy of $\pi_0$.
Each $E_n$ and therefore their sum obeys
\be
\lambda \di_\lambda E = - (d+2) E+ 2 \omega  \di_\omega E \; .
\ee
We thus see that if  a stationary point of $E$ exists, it must have $E=0$. This is not impossible {\em a priori} in our theory.
Due to the peculiar higher-derivative structure of the action, the galileon can violate the null energy condition without obvious pathologies in the low-energy effective theory \cite{NRT2}. This in principle allows for localized non-trivial field configurations with vanishing -- even negative -- total energy. 

Incidentally, eq. (\ref{En}) implies that there is another set
of zero modes, namely deformations described by eq. (\ref{rescale})
but restricted to $\lambda^2 \omega = 1$. Such a deformation
changes the overall energy by $1/\lambda^{d+2}$, but since the soliton
(if it exists) has vanishing $E$, so does its deformation.
In infinitesimal form ($\lambda = 1 + \epsilon$), the
deformation is
$\varphi_\epsilon = - 2 \epsilon \, \pi_0 + \epsilon \, \vec x \cdot \vec \nabla \pi_0.$
Thus our proof of the instability of the galileon soliton
could have made use of this zero-mode instead of the translational
zero-mode.

Let us close with a discussion of another class of derivatively coupled
theories for which essentially the same arguments apply,
\begin{eqnarray} \label{PofX}
S = \int d^{d+1} x \, P\big( (\partial \phi)^2 \big) \; , 
\end{eqnarray}
where $P$ is an arbitrary function. Such an effective theory describes for instance the low-energy dynamics of a superfluid \cite{son}, or, with more than one field, of ordinary fluids and solids \cite{DGNR} (our arguments also apply for multifield generalizations of (\ref{PofX})). It also describes more exotic systems, like the ghost condensate \cite{ghostcondensate1}, special cases of k-essence \cite{kessence}, and simpler ones like
a Goldstone boson non-linearly realizing a $U(1)$ symmetry, $\phi \to \phi+ c$.
Assuming there exists a soliton solution, one can see that $\lambda \omega = 1$ is the flat direction in this case, on top of the translations ones of course.
Provided there are spatial regions where small fluctuations
carry a positive energy, 
as they should far away from the supposed soliton, there should
also be regions where suitably chosen localized perturbations can destabilize the soliton.
Indeed it is obvious that the perturbations' quadratic energy is still of the form (\ref{quadratic_energy}), of course now with a different $Z^{ij}$.

Notice that most of the systems mentioned above as examples for eq.~(\ref{PofX}) spontaneously break Lorentz invariance. So the physical question in this case is whether there are soliton solutions in the broken phase. Yet our proof never makes use of Lorentz invariance, so it applies unaltered in the broken phase.  Perhaps more relevant is the worry for spontaneous breaking of time translations; a superfluid, the ghost condensate, and k-essence all break time translations, thus making the definition of energy in the broken phase  more subtle than usual. However there is a linear combination of time translations and  shift symmetry ($\phi \to \phi + c$) that is unbroken. The corresponding Noether charge is a perfectly good energy for excitations in the broken phase \cite{ghostcondensate1}, which our soliton solution should minimize, and which, apart from having a different `tensor' structure,  has the same scaling properties  as the original energy, thus lending itself to our proof. It should be emphasized however that the ghost condensate is degenerate at the lowest derivative level, so that the excitations' gradient energy is in fact dominated by higher-derivative terms, of the form $(\nabla^2 \pi)^2$, while the interactions are those given by eq.~(\ref{PofX}) and therefore have one derivative per field. This mismatch impairs the simple kinetic structure of the quadratic fluctuation Lagrangian \eqref{quadratic_energy}.  Therefore our proof does not hold for the ghost condensate. 

Finally let us comment on the quantum properties of a theory like eq.~(\ref{PofX}), or of its multi-field generalizations describing fluids and solids. 
Consider the structure of divergences in the one-loop effective action. 
Assuming for simplicity that $P$ is a generic function where all powers of $(\di \phi)^2$  are weighed by the same  scale $\Lambda$, we have
\be
\Gamma^{\rm 1-loop} \sim \sum_m \big[ \Lambda^4 + \Lambda^2 \di^2 +
\di^4 \log \Lambda^2/\di^2  \big] \bigg( \frac{(\di \phi)^2}{\Lambda^4} \bigg)^m  \; ,
\ee
where we cut off the UV divergences at $\Lambda$. The quartic divergence renormalizes by order one terms already present in $P((\di \phi)^2)$. The other contributions -- the quadratic divergence and the logarithmic one -- involve more derivatives per field. That is, apart from a trivial renormalization of the tree-level Lagrangian, here too quantum effects are negligible even for solutions where $(\di \phi)^2$ has large overall variations, as long as {\em derivatives} of $(\di \phi)^2$ are everywhere  small. The same is true at all loops as well. We even have experimental evidence that this is correct for certain systems -- we can easily compress a fluid to a fraction of its original volume ($\Delta(\di \phi)^2 \sim \Lambda^4$) without exiting the effective theory, provided we do so slowly enough ($\di \ll \Lambda$). 
Like for the galileon, if $\Lambda$ is really the only scale in the theory, our hypothetical soliton would have a size of order $\Lambda^{-1}$, and quantum effects would be large. We can bypass this problem by having very different scales in the theory and a see-saw mechanism, as discussed above. But this  requires a tuning, which here is not technically natural because, unlike for the galileon, the tree-level Lagrangian does get renormalized.
Our argument shows that, even allowing for such a tuning, there cannot be stable solitons in these systems \footnote{Localized, stationary vortices in a fluid do not conform to our definition of solitons: for them the velocity field is indeed time-indpendent -- yet it involves a time derivative of the canonical variables $\phi^I$ appearing in the Lagrangian. That is, for a vortex the $\phi^I$'s have a non-trivial time-dependence \cite{DGNR}.}.

{\em Acknowledgements.}
The authors would like to thank S.~Dubovsky, R.~Rattazzi and 
especially E.~Weinberg for very useful discussions and comments.
LH thanks NYU and IAS for hospitality.
This work is supported in part by the DOE (DE-FG02-92-ER40699)
and NASA ATP (09-ATP09-0049).  KH acknowledges support by funds provided by the University of Pennsylvania.


\end{document}